\documentstyle[sprocl]{article}

\def\be{\begin{equation}}
\def\ee{\end{equation}}
\def\bea{\begin{eqnarray}}
\def\eea{\end{eqnarray}}

\begin{document}

\title{Wigner Functions in Curved Spacetimes and Deformation Quantisation
of Constrained Systems}

\author{Frank Antonsen}

\address{Niels Bohr Institute}

\maketitle\abstracts{We first generalise the standard Wigner function to
Dirac fermions in curved spacetimes. Secondly, we turn to the Moyal 
quantisation of systems with constraints. Gravity is used as an example.}
  
\section{Curved Spacetime Wigner Function}
Probably the greatest unsolved problem of modern theoretical physics is the
interplay between quantum theory and gravity (as described by general
relativity).\\
The simplest ``subproblem'' is the study of quantum systems in a given, fixed
classical gravitational field, i.e., in a given curved background.\\
The Wigner function in flat space is
\be
	W(q,p) :=\int \bar{\psi}(q-\frac{1}{2}y)\psi(q+\frac{1}{2}y) e^{-iyp}
	\frac{d^ny}{(2\pi)^n}
\ee
in $n$ dimensions. The problem in curved spacetimes is the definition of
$q\pm\frac{1}{2}y$, since this in general doesn't make sense in non-flat 
spaces. In \cite{curved} I showed how to generalise $W$ to a Dirac spinor
in a given curved background. The Dirac equation for $\psi$ then induces the
following equation for $W$
\be
	\left[m+\gamma^\mu\left(e_\mu^ap_a+\frac{1}{2}i\nabla_\mu\right)\right]
	\hat{W} =-\frac{1}{2}\kappa\gamma^a\hat{X}_a\hat{W}
\ee
where $e_\mu^a$ is a vierbein and where $\hat{X}_a$ is an infinite order
differential operator involving the curvature tensor.\\
This allows one to find non-perturbative expressions for various macroscopic
quantities (i.e., quantum magneto-hydrodynamics). With this, a phasespace
interpretation of e.g. the conformal anomaly can be obtained, see \cite{curved}
for further details.

\section{Hamiltonian Systems with Constraints}
Gravitation itself is described by a set of constraints. The Hamiltonian itself
is nothing but a linear combination of constraints. This is in contrast to the
situation for other gauge fields (Maxwell or Yang-Mills), implying that we 
cannot simply import known results concerning gauge fields to gravitational
degrees of freedom.\\
Instead what we will do, is to perform a Moyal (or deformation) quantisation.
I.e., replace the classical Poisson brackets $\{\cdot,\cdot\}_{\rm PB}$ by
Moyal brackets $[\cdot,\cdot]_M$.
\bea
	[f,g]_M &=& i\hbar\{f,g\}_{\rm PB}+O(\hbar^2)\\
	&=& 2i f \sin\left(\frac{1}{2}\hbar\{\cdot,\cdot\}_{\rm PB}\right)g\\
	&=& f*g-g*f
\eea
Consequently, if we have a set of classical
constraints $\phi_a(q,p)$, satisfying $\{\phi_a,\phi_b\}_{\rm PB} = c_{ab}^c
\phi_c$ (i.e., being first class) we want to find a corresponding set of
quantum constraints $\Phi_a=...\hbar^{-1}\Phi_a^{(-1)} +\phi_a+\hbar
\Phi_a^{(1)}+...$ satisfying
\be
	[\Phi_a,\Phi_b]_M = i\hbar c_{ab}^c\Phi_c
\ee
It has been proven in \cite{deform} that classical second class constraints
(i.e., constraints satisfying $\{\phi_a,\phi_b\}_{\rm PB}\neq c_{ab}^c\phi_c$)
can be turned into quantum first class constraints by allowing a $\hbar^{-1}$
term in $\Phi_a$. If we cannot take $\Phi_a=\phi_a$ then we say we have an
anomaly. It was also proven in \cite{deform} how such anomalies could in
certain circumstances be ``lifted'', i.e., quantum constraints $\Phi_a$ did
exist satisfying the appropriate quantum (Moyal) constraint algebra. This is
the case if the anomaly is merely a central extension.\\
Classically, physical states are defined by the requirement $\forall a~:~\phi_a
=0$. In the standard Dirac quantisation scheme, this is interpreted as the
condition $\forall a~:~\hat{\phi}_a|\psi\rangle = 0$ picking out physical
states $|\psi\rangle$. In a deformation quantisation we must instead
introduce the BRST-like condition
\be
	[\Phi_a,W]_M^+ := 2\Phi_a\cos\left(\frac{1}{2}\hbar\{\cdot,\cdot\}_{\rm
	PB}\right)W = \Phi_a*W+W*\Phi_a=0
\ee
on the would-be physical Wigner functions $W$. In general, this will be an
infinite order differential equation.

\section{Gravity}
It turns out, that gravity in both the ADM approach (i.e., where the phasespace
is parametrised by the set of spatial 3-metrics and their conjugate momenta)
and in the Ashtekar variables approach (where one has a densitised dreibein
$E_a^i$ - a kind of electrical field - whose conjugate is a complex $SU(2)$
connection $A_i^a$ - the analogue of the Yang-Mills connection) is anomalous.
However, in the Ashtekar variables this anomaly is exceedingly simple
\be
	[{\cal H}(x),{\cal D}_i(x')]_M = i\hbar\{{\cal H}(x),{\cal D}_i(x')
	\}_{\rm PB} -9i\hbar^3\delta_{,i}(x,x')
\ee
being merely a central extension (a Schwinger term).  Here ${\cal H}= F_{ij}^a
E^i_bE^j_c\varepsilon_{bc}^{~~~a}$ is the
Hamiltonian constraint (in the obvious notation with $F_{ij}^a$ the field
strength tensor of $A_i^a$), and ${\cal D}_i=F_{ij}^aE^j_a$ the 
diffeomorphism one. The above Moyal bracket is the only anomalous one.\\
Furthermore, the equations for physical states become finite order, whereas
they become infinite order in the ADM-approach.
\bea
	0=[{\cal H},W]_M^+ &=& 2{\cal H}W-\frac{1}{2}\hbar^2\left(E_b^kE_v^l
	\epsilon_a^{~bc}\epsilon^a_{~ef}\frac{\delta^2W}{\delta E^k_e\delta
	E^l_f}-\right.\nonumber\\
	&&2\epsilon_a^{~bc}\left(-\delta^a_e(\delta^k_i\partial_j-\delta^k_j
	\partial_i)+\epsilon^a_{~pq}(\delta^p_e\delta^k_iA^q_j
	+\delta^q_e\delta^k_jA_i^p)\right)\times\nonumber\\
	&&(\delta^i_l\delta^b_fE^j_c
	+\delta^j_l\delta^c_fE^i_b)\frac{\delta^2W}{\delta E_e^k\delta A^f_l}
	+\nonumber\\
	&&\left.\epsilon_a^{~bc}F_{ij}^a\frac{\delta^2W}{\delta A_i^b
	\delta A_j^c}\right)+\frac{5}{4}\hbar^4\epsilon_{bc}^{~~a}
	\epsilon_a^{~ef}\frac{\delta^4W}{\delta E^k_e\delta E^l_f\delta A^e_k
	\delta A^f_l}\\
	0=[{\cal D}_i,W]_M^+ &=&2{\cal D}_iW-\frac{1}{2}\hbar^2\left(
	\epsilon^a_{~ef}E^j_a\frac{\delta^2W}{\delta E_e^i\delta E^j_f}-
	\right.\nonumber\\
	&&2\left(-\delta^a_e(\delta^k_i\partial_j-\delta^k_j\delta_i)
	+\epsilon^a_{~mn}(\delta^m_e\delta^k_iA^m_j+\delta^n_e\delta^k_j
	A^m_i)\right)\frac{\delta^2W}{\delta E^k_e\delta A^a_j}\nonumber\\
	&&\\
	0=[{\cal G}_a,W]_M^+ &=& 2{\cal G}_aW + \frac{1}{4}i\hbar^2\delta^j_k
	\epsilon^c_{~ab}\frac{\delta^2W}{\delta A^c_j\delta A^k_b}
\eea
with ${\cal G}_a=D_iE^i_a$ the Gauss constraint.

\end{document}